\documentclass[3p,times]{elsarticle}

\usepackage{ecrc}
\usepackage{color}

\volume{00}

\firstpage{1}

\journalname{Physics Procedia}

\runauth{}

\jid{phpro}

\jnltitlelogo{Physics Procedia}

\usepackage{amssymb}

\usepackage[figuresright]{rotating}

\begin{document}

\begin{frontmatter}



\dochead{}

\title{{\small12$^{\rm th}$ International Conference on Muon Spin Rotation, Relaxation and Resonance}\\
Magnetic order and frustrated dynamics in Li(Ni$_{0.8}$Co$_{0.1}$Mn$_{0.1}$)O$_2$:\\a study by $\mu^{+}$SR and
SQUID magnetometry}


\author[UU1]{J.~Magnus~Wikberg}
\author[ETH,PSI]{Martin~M\aa{}nsson\corref{cor1}},
\cortext[cor1]{Corresponding author. Tel.: +41-(0)56-310-5534 ; fax: +41-(0)44-633-1282}
\ead{mansson@phys.ethz.ch}
\author[UU2]{Mohammed~Dahbi}
\author[TCRDL]{Kazuya~Kamazawa}
\author[TCRDL]{Jun~Sugiyama}

\address[UU1]{Department of Engineering Science, Uppsala University, Box 534, SE-751 21 Uppsala, Sweden}
\address[ETH]{Laboratory for Solid state physics, ETH Z\"{u}rich, CH-8093 Z\"{u}rich, Switzerland}
\address[PSI]{Laboratory for Neutron Scattering, Paul Scherrer Institute, CH-5232 Villigen PSI, Switzerland}
\address[UU2]{Department of Materials Chemistry, Uppsala University, Box 538, SE-751 21 Uppsala, Sweden}
\address[TCRDL]{Toyota Central Research and Development Labs. Inc., Nagakute, Aichi 480-1192, Japan}

\begin{abstract}
Recently, the mixed transition metal oxides of the form Li(Ni$_{1-y-z}$Co$_{y}$Mn$_{z}$)O$_{2}$, have become the center of attention as promising candidates for novel battery material. These materials have also revealed very interesting magnetic properties due to the alternate stacking of planes of metal oxides on a 2D triangular lattice and the Li-layers. The title compound, Li(Ni$_{0.8}$Co$_{0.1}$Mn$_{0.1}$)O$_{2}$, has been investigated by both magnetometry and measurements and $\mu^{+}$SR. We find the evolution of localized magnetic moments with decreasing temperature below 70~K. The magnetic ground state ($T=2$~K) is, however, shown to be a frustrated system in 3D, followed by a transition into a possible 2D spin-glass above 22~K. With further increasing temperature the compound show the presence of remaining correlations with increasing effective dimensionality all the way up to the ferrimagnetic transition at $T_{\rm C}=70$~K.
\end{abstract}

\begin{keyword}
Muon-spin relaxation/rotation ($\mu^{+}$SR) \sep frustrated magnetism \sep battery materials


\end{keyword}

\end{frontmatter}


\section{Introduction}
Historically the triangular lattice system has served as a playground for new ideas about various unconventional phases of frustrated antiferromagnets \cite{Anderson}. In this context, the layered cobalt dioxides have been in the center of attention for many years \cite{Sugiyama1,Mukai_LCO}. The two main driving forces have been the electrochemical importance of the LiCoO$_{2}$ compound as an electrode in Li-ion batteries \cite{Mizushima}, and by the unconventional superconductivity discovered in the Na$_{0.35}$CoO$_{2}\cdot$1.3H$_{2}$O compound \cite{Takada}. Indeed, one of the main obstacles for the maturity of electric cars is the development of a high-capacity, cheap and safe rechargeable battery. The most widely used cathode material is by far LiCoO$_{2}$ \cite{Sugiyama_DLi}, however, cobalt is very expensive and there is a strong driving force to find new cheaper and environmental friendly cathode materials. One of the big problems for battery application is the structural distortions that occur when large amount of Li is removed \cite{Ven,Seguin}. Some efforts have been made to minimize these effects e.g. by the combination of Co and Ni into a LiNi$_{1-x}$Co$_{x}$O$_{2}$ solid solution. However, this compound suffers from strong intermixing of Ni into the Li layers for the highly delithiated state \cite{Chebiam}. Recently, the mixed transition metal oxides (MTMO's) of the form LiNi$_{1-y-z}$Co$_{y}$Mn$_{z}$O$_{2}$ have become the center of attention \cite{Yoshio,Choi}. In particular has the LiNi$_{1/3}$Co$_{1/3}$Mn$_{1/3}$O$_{2}$ compound \cite{Ohzuku} been put forward as one of the most promising candidates as a novel battery electrode material. The structure of these compounds is the same as for the fundamental LiCoO$_{2}$ i.e. a rhomohedral lattice (space group $R\overline{3}mH$) where Ni/Co/MnO$_{2}$ planes are stacked between nonmagnetic Li layers along the $c$-axis [see Fig.~1(a)]. In similarity with LiNiO$_{2}$ some intermixing of Ni in the Li layers is to be expected. However, in this particular case, this actually helps to stabilize the structure and improve capacity retention \cite{Dahbia}. These materials have also revealed very interesting and complex magnetic properties partially due to the 2D triangular lattice created by the transition metal ions \cite{Wikberg_PRB}.
\begin{figure}
\begin{center}
\includegraphics[keepaspectratio=true,width=160 mm]{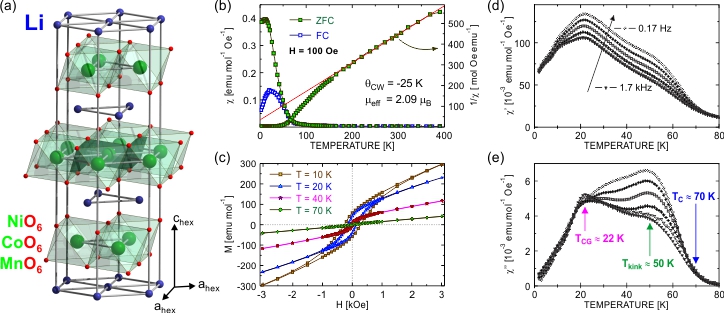}
\caption{\label{fig:1}(a) The crystal structure of Li(Ni$_{0.8}$Co$_{0.1}$Mn$_{0.1}$)O$_2$ showing the Li-ion planes in between the MO$_{6}$ (M~=~Ni, Co, Mn) planes. Note that some Ni ($<1\%$) is expected to be located within the Li layers. (b) Temperature dependence of DC-susceptibility [$\chi(T)$] at $H=100$~Oe for ZFC (open symbols) and FC (filled symbols) protocol. The inverse susceptibility [$\chi^{-1}(T)$], derived from the FC data, with a Curie--Weiss fit (solid red line) yielding $\mu_{\rm eff}=2.09\mu_{\rm B}$ and $\theta_{\rm CW}=-25$~K. In addition, a ferrimagnetic like transition is seen around $T_{\rm C}=70$~K. (c) Magnetization ($M$) as a function of field ($H$) at $T=$~10, 20, 40, and 70 K. (d-e) AC-susceptibility showing $\chi'(T)$ and $\chi''(T)$, respectively, acquired at the frequencies $f=$~0.17, 1.7, 17, 170, 1000, and 1700 Hz, using an AC-field $H_{\rm AC}=2$~Oe.}
  \label{fig:structure}
\end{center}
\end{figure}

We here present a muon-spin relaxation/rotation ($\mu^{+}$SR) investigations of the microscopic magnetic properties of the MTMO compound Li(Ni$_{0.8}$Co$_{0.1}$Mn$_{0.1}$)O$_2$. From our AC- and DC-magnetometry measurements (see Fig.~1(b-e) and \cite{Wikberg_JAP}) have been found to be a percolating spin system interacting via AF [negative Curie-Weiss temperature, $\theta_{\rm CW}=-25$~K, see Fig.~1(b)] and ferromagnetic (FM) exchange interactions [$T_{\rm C}=70$~K and the apparent hysteresis in Fig.~1(c)] of different strength. Hereby, a quasi long-range ferrimagnetic order is induced, with no translational symmetry of the interactions. On cooling, signs of a 2D spin glass (SG) is found [c.f. field-cooled (FC) and zero-field-cooled (ZFC) protocols in Fig.~1(b) as well as frequency dependence in Fig.~1(d-e)], followed by a completely frustrated system in 3D at the lowest temperature.

\begin{figure}
\begin{center}
\includegraphics[keepaspectratio=true,width=140 mm]{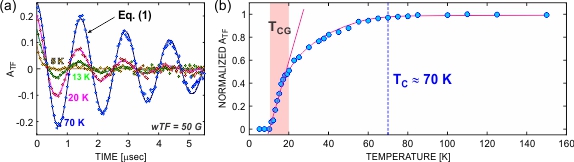}
\caption{\label{fig:1}(a) Weak transverse-field (wTF~=~50~G) $\mu^{+}$SR time spectra obtained at different temperatures. Solid lines are fits to Eq.~(1). (b) Fit results showing the temperature dependence of the wTF asymmetry ($A_{\rm TF}$). Two critical temperature regions are visible, $T_{\rm CG}\approx10-20$~K and $T_{\rm C}=70$~K.}
\end{center}
\end{figure}

\section{\label{sec:E}Experimental Details}
Approximately 1 g of a Li(Ni$_{0.8}$Co$_{0.1}$Mn$_{0.1}$)O$_2$ powder sample was placed in a small envelope made of very thin Al-coated Mylar tape and then attached to a low-background, fork-type, sample holder. In order to make certain that the muons stopped primarily inside the sample, we ensured that the side facing the muon beamline was only covered by a single layer of Mylar tape. Subsequently, $\mu^+$SR spectra were measured at the Swiss Muon Source (S$\mu$S), Paul Scherrer Institut, Villigen, Switzerland. By using the muon beamline $\pi$E1 and the Dolly spectrometer, zero-field (ZF) and weak transverse-field (wTF) spectra were collected for 2~K~$\leq{}T\leq$~150~K. The experimental setup and techniques were described in detail elsewhere \cite{Kalvius}.

\section{\label{sec:E}Results \& Discussion}
As shown in Fig.~2(a), weak transverse-field (wTF~=~50~G) $\mu^{+}$SR spectra were obtained as a function of temperature. Below $T=10$~K a total suppression of the externally applied field is seen as annihilation of the wTF asymmetry ($A_{\rm TF}$). The data is found to be well fitted by a combination of an exponentially relaxing cosine oscillation and a fast relaxing component:
\begin{eqnarray}
 A_0 \, P_{\rm TF}(t) = A_{\rm TF}\cos(2\pi f_{\rm TF}\cdot{}t+\phi_{\rm TF})\cdot{}e^{-\lambda_{\rm TF} t} + A_{\rm fast}\cdot{}e^{-\lambda_{\rm fast} t~}~,
\label{eq:ZFfit}
\end{eqnarray}
From such fits, the temperature dependence of $A_{\rm TF}$ can be extracted and a clearer picture of the transition is obtained [Fig.~2(b)]. Moving from low to high temperature: at $T=10$~K an evident transition occurs and $A_{\rm TF}$ is linearly recovered up until $T=20$~K. From the AC-susceptibility data shown above, this temperature is identified as the 3D cluster glass (CG) transition ($T_{\rm CG}$) \cite{Wikberg_JAP}. However, as seen from the pink-shaded area in Fig.~2(b), only 50\% of the total $A_{\rm TF}$ is recovered in this transition. The remainder is thereafter more slowly restored within the ferrimagnetic quasi long-range ordered phase \cite{Wikberg_JAP} and finally reaches its full value at $T_{\rm C}=70$~K.

To obtain a more detailed view of the nature of the magnetic order and transitions in this compound, also zero-field (ZF) data was recorded. At the lowest temperature ($T=2$~K), the $\mu^{+}$SR time spectrum show no indication of spontaneous muon precession, but rather only a very fast decaying signal, indicating either a wide field-distribution (typical for spin-glasses) or the presence of dynamical correlations. To obtain an adequate distinction between these two cases, a longitudinal-field (LF) experiment would be needed, where the static contribution could be decoupled. However, from the information obtained in our magnetometry data, the static case seems more likely. The ZF data was found to be well fitted to the sum of a very fast exponentially relaxing signal, a slower decaying stretched-exponential \cite{Phillips} function and a powder average tail component:
\begin{eqnarray}
 A_0 \, P_{\rm ZF}(t) = A_{\rm fast}\cdot{}e^{-\lambda_{\rm fast} t~} + A_{\rm slow}\cdot{}e^{[(-\lambda_{\rm slow} t)^{n}]} + A_{\rm tail}\cdot{}e^{-\lambda_{\rm tail} t~},
\label{eq:ZFfit}
\end{eqnarray}
From the temperature dependent measurements displayed in Fig.~3(b), the evolution of the two components is evident. At low temperature the fast relaxing component is dominating. However, when increasing the temperature, the fast decaying signal quickly fades and is almost gone at $T=16$~K, showing that it is connected to the SG transition. Around $T=20$~K, the slowly relaxing component has completely taken over. This kind of stretched exponential relaxation is found to appear in a range of different systems. However, they are commonly connected to spin-glasses and systems containing magnetic frustration i.e. like the title compound. The critical exponent $n$ is known to depend on the effective dimensionality ($d$) of the system above (but in the vicinity of) a transition according to $n=\frac{d}{d+1}$. In Li(Ni$_{0.8}$Co$_{0.1}$Mn$_{0.1}$)O$_2$, $n$ display a strong temperature evolution. At $T=70$~K, $n=2$ i.e. a gaussian relaxation (or possible Kubo-Toyabe behavior) indicating the presence of random moments. With lowering of the temperature, $n$ drastically starts to decrease around $T=40$~K and close to 20 K, $n\leq1/2$, indicating a strongly decreased dimensionality (2D or 1D) just above $T_{\rm CG}$. This could be a signature for the onset of the previously reported glassy transition in 2D and 3D, respectively \cite{Wikberg_JAP} arising from the structurally induced frustration on the 2D triangular lattice. It is also tempting to connect the onset of decrease in $n$ to the kink found in the AC-susceptibility data [seee Fig.~1(e)]. However, further measurements including a more detailed temperature dependence is needed in order to clarify this matter more robustly.

\begin{figure}
\begin{center}
\includegraphics[keepaspectratio=true,width=100 mm]{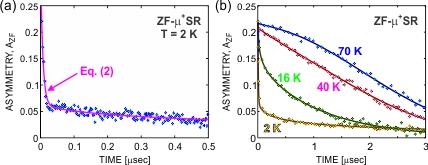}
\caption{\label{fig:1}(a) Zero-field (ZF) $\mu^{+}$SR time spectra obtained at lowest temperature ($T=2$~K), showing the absence of muon precessions and instead a very fast relaxing component. (b) Temperature evolution of the ZF data in a longer time domain. Solid lines are all fits to Eq.~(2).}
\end{center}
\end{figure}

\section{\label{sec:E}Summary}
We present a study of local magnetic properties in the mixed transition metal oxide, Li(Ni$_{0.8}$Co$_{0.1}$Mn$_{0.1}$)O$_2$. wTF data show how the localized magnetic moments are gradually building up in steps over an extended $T$-range. Moreover, the ZF results show a spin glass state at the lowest temperatures with a transition (on heating) from a 3D to a 2D spin glass state around 20 K. The absence of long-range magnetic order, even at the lowest temperature measured, could be directly connected to magnetic frustration of the 2D triangular lattice. With further increasing temperature the compound show the presence of remaining correlations with increasing effective dimensionality up to the ferrimagnetic transition at $T_{\rm C}=70$~K. The $\mu^{+}$SR results gives direct support for conclusions made in our previously published results \cite{Wikberg_JAP}.

\paragraph{\textbf{Acknowledgments}}\

This work was performed using the \textbf{Dolly} muon spectrometer at the Swiss Muon Source (S$\mu$S) of the Paul Scherrer Institut (PSI), Villigen, Switzerland and we are thankful to the instrument staff for their support. All the $\mu^{+}$SR data was fitted using \texttt{musrfit} \cite{musrfit} and the images involving crystal structure were made using the DIAMOND software. This research was financially supported by the Swedish Research Council (VR), Swedish Energy Agency, Knut and Alice Wallenberg Foundation (KAW), the Swiss National Science Foundation (through Project 6, NCCR MaNEP), and Toyota Central Research \& Development Labs. Inc.

\paragraph{\textbf{References}}\





\bibliographystyle{elsarticle-num}



\end{document}